\documentclass[prb,floatfix,twocolumn,showpacs,amsmath,amssymb]{revtex4}
\usepackage{graphicx}
\usepackage{dcolumn}
\usepackage{bm}
\usepackage{epsfig,psfrag,amsmath,amssymb,float,psfrag}
\input{epsf}

\newcommand{\bc}{\begin{center}}
\newcommand{\ec}{\end{center}}
\newcommand{\be}{\begin{equation}}
\newcommand{\ee}{\end{equation}}
\newcommand{\beqn}{\begin{eqnarray}}
\newcommand{\eeqn}{\end{eqnarray}}

\def\1.2{\frac{1}{2}}

\begin{document}

\title{Scaling of Entanglement Entropy in the Random Singlet Phase}


\author{Nicolas Laflorencie}

\affiliation{Department of Physics \& Astronomy, University of British Columbia,
Vancouver, B.C., Canada, V6T 1Z1}

\date{\today}

\begin{abstract}
We present numerical evidences for the logarithmic scaling of the
entanglement entropy in critical random spin chains. Very large scale exact
diagonalizations performed at the critical XX point up to
$L=2000$ spins $\frac{1}{2}$ lead to a perfect agreement with recent
real-space renormalization-group predictions of Refael and Moore
[Phys. Rev. Lett. {\bf 93}, 260602 (2004)] for the logarithmic
scaling of the entanglement entropy in the Random Singlet Phase with an effective
central charge ${\tilde{c}}=c\times \ln 2$. Moreover we provide the
first visual proof of the existence the Random Singlet Phase with the
help of quantum entanglement concept.
\end{abstract}

\maketitle
The study of quantum phase transitions through quantum entanglement concepts 
provides a new way to understand strongly correlated systems near criticality.
In one dimensional systems, such as quantum spin chains, 
entanglement estimators exhibit universal features close to a critical point~\cite{Nature,Vidal03}.
One of this estimator is the {\it entanglement entropy} of a subsystem A with respect to a subsystem B.
Defined as the Von Neumann entropy of the reduced density matrix for either subsystem
\be
{\mathcal{S}}=-Tr {\hat{\rho}}_A \ln {\hat{\rho}}_A=-Tr {\hat{\rho}}_B \ln {\hat{\rho}}_B,
\ee
this quantity displays very interesting scaling behavior for conformally invariant critical theories in one 
dimension (1D). Indeed, as  shown first by Holzey, Larsen and Wilczek~\cite{HLW94} in the context of geometric entropy related to black hole physics, 
the entanglement entropy of a subsystem of length $x$ embedded in an infinite system is expected to scale like
\be
{\mathcal{S}}(x)\sim\frac{c}{3}\ln x.
\label{eq:1}
\ee
The number $c$ is the so-called central charge which is, for instance,
for the critical XXZ spin-$\1.2$ chain $c_{\rm XXZ}=1$ or 
for the spin-$\1.2$ Ising chain in transverse field at criticality
$c_{\rm Ising}=1/2$.
This result [Eq.~(\ref{eq:1})] has been verified numerically~\cite{Vidal03,Vidal04} as well as analytically in Ref.~\cite{Korepin04} where 
some simple connections have
been established between thermodynamic entropy and entanglement entropy.   
An important extension to critical and non-critical systems with finite size,
finite temperature and different boundary conditions has been achieved by
Calabrese and Cardy~\cite{Cardy04}. They showed for instance 
that for critical systems of finite size $L$ with periodic boundary
conditions, 
Eq.~(\ref{eq:1}) should be replaced by
\begin{equation}
{\mathcal{S}}(L,x)=\frac{c}{3}\ln \left(\frac{L}{\pi}\sin(\frac{\pi x}{L})\right)+{\rm{s_1}},
\label{eq:2}
\end{equation}
where $s_1$ is a constant related to the UV cut-off.

\begin{figure}[!ht]
\bc
\includegraphics[width=\columnwidth]{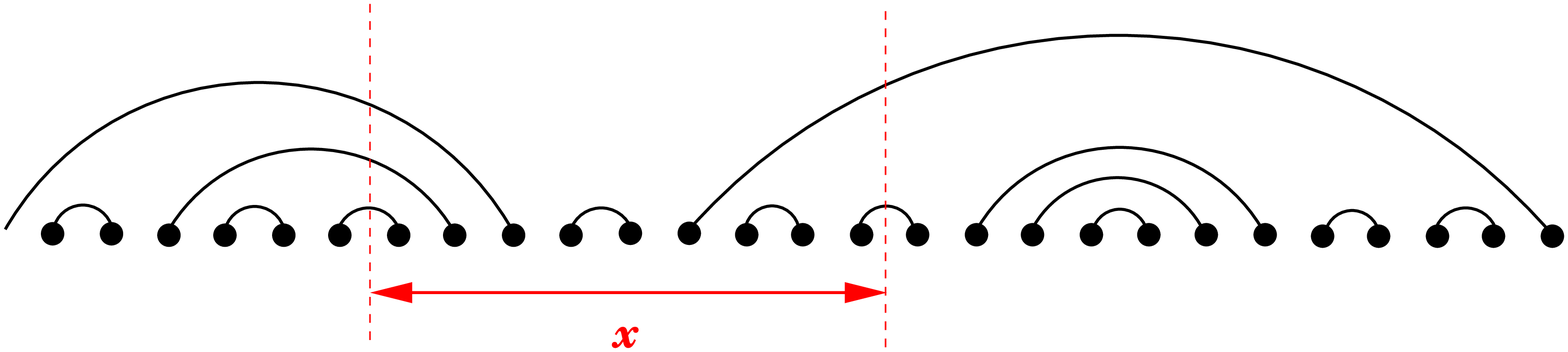}
\caption{Schematic picture for the entanglement entropy of a subsystem
  of length $x$ in the random singlet phase. The entanglement is just
  due to the singlets connecting the subsystem with the rest of the
  chain. In this picture, ${\mathcal{S}}(x)=5\times \ln 2$.}
\label{fig:1}
\ec
\end{figure}

Although such a logarithmic scaling of the entanglement entropy seems closely related to the conformal invariance of the critical system, it has been shown recently 
by Refael and Moore~\cite{Refael04} that such a critical scaling is also expected for some random critical points. Indeed, using an analytic real-space renormalization-group (RSRG) approach, they have shown that random critical spin chains display similar features that clean ones
with an {\it{effective central charge}} ${\tilde{c}}=c\times \ln 2$ so that
\be
{\mathcal{S}}(x)=\frac{{\tilde{c}}}{3}\ln x + {\rm{constant}}.
\label{eq:RM}
\ee
This surprising results can be derived using the RSRG method
introduced by Ma, Dasgupta and Hu~\cite{Ma} several years ago to
study random spin chains. As a result, any amount of randomness introduced
as a perturbation in a clean XXZ critical spin-$\1.2$ chain is
relevant~\cite{note1} and drives the system to the so-called Random Singlet Phase
(RSP)~\cite{Fisher}, associated with an infinite randomness fixed
point (IRFP) for the RSRG transformation~\cite{Fisher}.
The RSP can be depicted as a collection of singlet bonds of arbitrary length (see
Fig.~\ref{fig:1}). 
%
Then, utilizing the very simple result that the
entanglement entropy of a spin $\1.2$ involved in a singlet with its
partner is $\ln 2$, in the RSP the entanglement of a segment with the
rest of the system is just
given by  $\ln 2$ times the number of singlets which cross the boundary of the
segment, as depicted in Fig.~\ref{fig:1}.
Using this fact as well as an accurate RSRG
calculation, Refael and Moore have then been able to determine
precisely that the number of singlets connecting a segment of size $x$
with the rest of the system is $(1/3) \ln x$, leading to the formula
(\ref{eq:RM}). 
The purpose of this communication is to
investigate numerically the entanglement of the RSP and compare
the RSRG prediction [Eq.~(\ref{eq:RM})] with exact computations.

{\it Exact computation of the entanglement entropy ---}
In order to compute the entanglement entropy of a subsystem, one needs to calculate the corresponding reduced density matrix. 
For general XXZ spin chains governed by
\be
{\cal{H_{\rm{XXZ}}}}=J\sum_{j}\left[\frac{1}{2}\left(S_{j}^{+}S_{j+1}^{-}+S_{j}^{-}S_{j+1}^{+}\right)+\Delta S_{j}^{z}S_{j+1}^{z}\right],
\ee
the non-critical regime (achieved if $|\Delta|>1$) can be investigated
using the corner transfer matrices of the corresponding
two-dimensional (2D) classical problem~\cite{Baxter,Peschel1}. On the
other hand, along the critical line ($-1\le \Delta\le 1$), an 
analytical computation of ${\mathcal{S}}(x)$ is more difficult and conformal field theory (CFT) tools are then required~\cite{Cardy04}. Another alternative consists in performing numerical 
exact diagonalizations (ED) of finite lengths spin chains, but it is limited to $L_{\rm{max}}\simeq 40$ spins $\frac{1}{2}$ when $\Delta\ne 0$~\cite{Andreas}. 
Nevertheless, the XX point $\Delta=0$ is special because the spin Hamiltonian can be rewritten using the Jordan-Wigner transformation as a free-fermions 
model
\be
{\cal{H}}_{\rm{XX}}=\frac{J}{2}\sum_{j}\left[c_{j}^{\dagger}c_{j+1}+c_{j+1}^{\dagger}c_{j}\right]
\label{eq:6}
\ee
for which the density matrix can be expressed as the exponential of a free-fermion operator~\cite{Chung01}. It turns out that the reduced density matrix is completely determined by the $x\times x$ correlation matrix
${\mathcal{C}}(x)$, defined by
\begin{equation}
\label{eq:C}
{\mathcal{C}}(x)~=~\begin{pmatrix}\langle c^{\dagger}_{1}c_1\rangle & \langle c^{\dagger}_{1}c_2\rangle  & \cdots &  \langle c^{\dagger}_{1}c_x\rangle \\
\langle c^{\dagger}_{2}c_1\rangle & \langle c^{\dagger}_{2}c_2\rangle & \ddots & \vdots\\
\vdots &  & \ddots &  \\
 & &  & \langle c^{\dagger}_{x}c_x\rangle
\end{pmatrix}
.
\end{equation}
The matrix elements ${\mathcal{C}}_{ij}=\langle c^{\dagger}_{i}c_j\rangle $ can be calculated either numerically by diagonalizing the free-fermion Hamiltonian 
in momentum space or analytically in some special cases~\cite{Peschel2}.
The entanglement entropy of a subsystem of size $x$ embedded in a
larger system is then given by
\be
{\mathcal{S}}(x)=-\sum_{k}\left[\lambda_k \ln \lambda_k +(1-\lambda_k) \ln
(1-\lambda_k)\right],
\label{eq:S}
\ee
where the $\lambda_k$ are the eigenvalues of ${\mathcal{C}}(x)$.

Let us now concentrate on the disordered XX spin-$\1.2$ chain, governed by
the random hopping Hamiltonian on a periodic ring of length $L$
\begin{eqnarray}
{\cal{H}}_{\rm{XX}}=\sum_{j=1}^{L-1}J_j\left[c_{j}^{\dagger}c_{j+1}+c_{j+1}^{\dagger}c_{j}\right]
\nonumber \\
+J_L\exp(i\pi{\cal{N}})(c_{L}^{\dagger}c_{1}+c_{1}^{\dagger}c_{L})\
\end{eqnarray}
where $J_j$ are positive random numbers chosen in a flat uniform
distribution within the interval $[0,1]$~\cite{note2}, and the second
term in the right hand side ensures that periodic boundary conditions
are imposed in the spin problem. The total number of
fermions is ${\cal N}=L/2$ in the ground-state (GS).
The way to diagonalize ${\cal{H}}_{\rm{XX}}$ is straightforward and
has already been explained by several authors~\cite{LSM61,Henelius98}.
As a check, we have first computed the entanglement entropy (\ref{eq:S}) for
clean systems (i.e. $J_i$ is a constant) 
of total sizes $L=500$ and $L=2000$. Technically, this
only involves computing the elements $\langle
c_{i}^{\dagger}c_{j}\rangle$ by diagonalizing the free-fermions
Hamiltonian (\ref{eq:6}), and
then one needs to diagonalize $\cal C$ [Eq.~(\ref{eq:C})] 
using standard linear algebra
routines~\cite{LAR}.
 \begin{figure}[!ht]
\bc
\psfrag{A}{\tiny{$0.8595+\frac{\ln 2}{3}\ln x$}}
\psfrag{Y}{\tiny{$0.72602+\frac{1}{3}\ln \left(\frac{2000}{\pi}\sin(\frac{\pi x}{2000})\right)$}}
\psfrag{Z}{\tiny{$0.72587+\frac{1}{3}\ln \left(\frac{500}{\pi}\sin(\frac{\pi x}{500})\right)$}}
\psfrag{E}{${\mathcal{S}}(L,x)$}
\psfrag{T}{$x$}
\includegraphics[width=\columnwidth,height=6.5cm,clip]{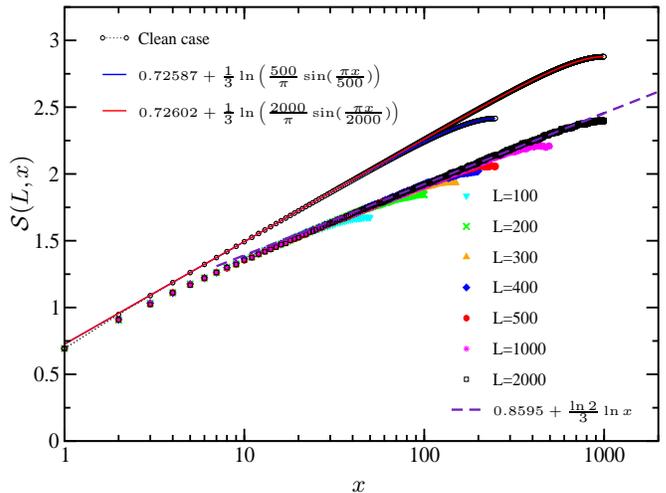}
\caption{(color online) Entanglement entropy of a subsystem of size
  $x$ embedded in a closed ring of size $L$, shown vs $x$ in a
  log-linear plot. Numerical results obtained by exact
  diagonalizations performed at the XX point. 
For clean non random systems with $L=500$ and $L=2000$ (open circles), ${\mathcal{S}}(x)$ is perfectly described by Eq.~(\ref{eq:2}) (red and blue curves).
The data for random systems  have been averaged over $10^4$ samples for $L=500,~1000,~2000$ and $2\times 10^4$ samples for $100\le L \le 400$.
The expression $0.8595+\frac{\ln 2}{3}\ln x$ (dashed line) fits the data in the regime where finite size effects are absent.}
\label{fig:2}
\ec
\end{figure}
The results are shown in Fig.~\ref{fig:2} where we can see that ${\mathcal{S}}(L,x)$
is perfectly described by the CFT prediction Eq.~(\ref{eq:2}).
Note also that the constant term is found to be $s_1\simeq 0.726$, in
excellent agreement with the recent analytical prediction of Jin and
Korepin~\cite{Korepin2}.

For the random case, the same technique has been used but a bigger
computational effort was necessary to average over a large number of
independent random samples. Practically the number of samples used was
$2\times 10^4$
for $L=100,~200,~300,~400$ and $10^4$ for $L=500,~1000,~2000$ which
required 2000 hours of CPU computational time.
The results for the disorder averaged entanglement entropy are shown
in Fig.~\ref{fig:2}
When the subsystem size $x$ is large enough (typically $x>20$),
the expression (\ref{eq:RM}) derived by Refael and Moore describes
perfectly the behavior of the disorder average entanglement entropy,
i.e. a logarithmic scaling with an effective central charge 
${\tilde{c}}=\ln 2$.
One can notice that when the subsystem size approaches $L/2$ some
finite size effects are visible, as it is the case in clean systems.

{\it Signature of the Random Singlet Phase ---}
The very good agreement found between exact numerical diagonalizations
and RSRG calculations for the entanglement properties in the RSP is a
new proof in favor of the random singlet nature of the GS, 
also supported by recent neutron
scattering experiments performed on the disordered spin chain compound 
BaCu$_2$(Si$_{0.5}$Ge$_{0.5}$)$_2$O$_7$~\cite{Masuda04}.
Another way to get more insight on these long distance effective singlets in
the GS consists in looking at the probability distribution 
of the entanglement entropy.
Indeed, since each singlet
is expected to contribute as a $\ln 2$ in the entanglement entropy, we
can focus on the probability distribution of ${\cal{S}}/\ln 2$ for a
given subsystem embedded in a larger system. In order to get a correct
statistical picture for the typical behavior of this random singlets
formation, one needs a huge number of disordered samples. We chose 
to study $10^5$ independent realizations. The price to pay is that
not too large systems can then be diagonalized. Nevertheless,
only focusing 
on $L=100$ spins is enough to get good insights on the RSP. Indeed,
instead of increasing the system size to achieve the physics of the
RSP, according to the disorder induced crossover phenomena observed
for the RSRG flow~\cite{comment03} one can rather keep $L$ fixed and 
increase the disorder strength to get closer to the IRFP
and therefore deeper in the RSP.
Let us thus consider strong disorder distributions for the couplings
$J_i$, like
\be
{\cal{P}}(J)=\frac{1}{\delta}J^{-1+\delta^{-1}},
\ee
parametrized by a disorder strength $\delta\ge 1$. This distribution
is quite natural to mimic strong disorder effects since at the IRFP, the fixed
point distribution for the random couplings is achieved for $\delta\to
\infty$.

 \begin{figure}
\bc
\psfrag{A}{{(c) $\delta=5$}}
\psfrag{B}{{(d) $\delta=10$}}
\psfrag{C}{{(b) $\delta=2.5$}}
\psfrag{D}{{(a) $\delta=1$}}
\psfrag{E}{\Large{${{\mathcal{S}}}/{\ln 2}$}}
\psfrag{P}{{\Large{$P^{{\rm{even}},{\rm{odd}}}$}}}
\includegraphics[width=\columnwidth,clip]{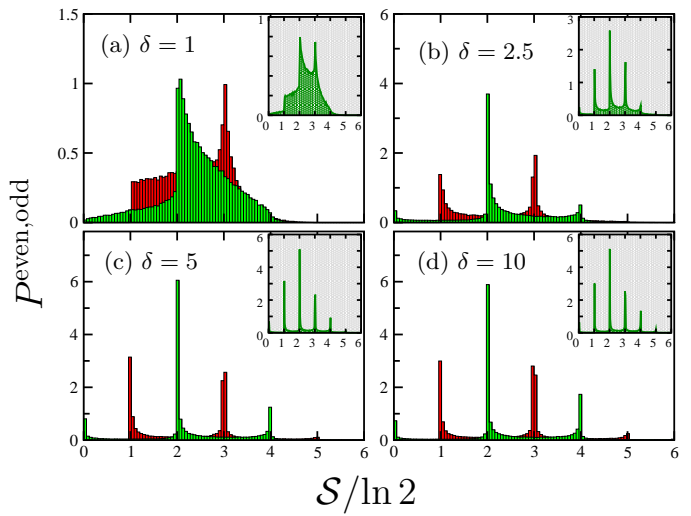}
\caption{(color online) Probability distribution for the entanglement
  entropy in the random singlet phase for the disordered XX
  spin-$\1.2$ chain obtained by exact diagonalizations on
  $L=100$ spin chains. Green histograms correspond to an even subsystem
  size with $50$ sites ($P^{\rm{even}}=P({\cal{S}}(100,50)/\ln 2)$) and
  red histograms correspond to an odd subsystem with $49$ sites ($P^{\rm{odd}}=P({\cal{S}}(100,49)/\ln 2)$).
  For each disorder strength, $\delta=1$ (a), $\delta=2.5$ (b),
  $\delta=5$ (c), $\delta=10$ (d), $10^5$ different random samples
  have been diagonalized. The insets show the combined distributions
  $P^{\rm{even}}+P^{\rm{odd}}$. Note that all the distribution
  functions have been normalized to unity.}

\label{fig:3}
\ec
\end{figure}
I order to minimize the finite size effects, we consider half of
the chain as a subsystem and compute ${\mathcal{S}}(L,L/2)$ for each
sample. Nevertheless, in order to get a good understanding, it is
important to notice that the parity of $L/2$ is crucial. Indeed if
$L/2$ is odd only an odd number of random singlets can connect both
subsystems whereas if $L/2$ is even, the number of cut singlets will
be even, none singlet being also a possibility. 
This fact is actually clearly visible in Figs.~\ref{fig:3} where we
have plotted the probability distributions
$P^{\rm{even}}=P({\mathcal{S}}(100,50)/\ln 2)$ (green histogram) as well as
$P^{\rm{odd}}=P({\mathcal{S}}(100,49)/\ln 2)$ (red histogram), for
$\delta=1,~2.5,~5,~10$.
Whereas for $\delta=1$ (Fig.~\ref{fig:3}(a)) $P^{{\rm{even}}}$ ($P^{{\rm{odd}}}$) displays an
integer-peaks structure, signature of
the RSP, only for ${\cal{S}}/\ln 2=2$ (${\cal{S}}/\ln 2=3$) and that a
non-negligible statistical weight lies between for non-integer values, 
when the disorder strength increases, the integer-peaks structure becomes
more and more pronounced as visible in Figs.~\ref{fig:3}(b-d).
The combined distributions $P^{\rm{even}}+P^{\rm{odd}}$ are also plotted
in the insets of Figs.~\ref{fig:3}.
Thanks to the entanglement entropy, we provide a clear visual proof
for the RSP.

{\it Discussion and conclusion ---}
Non disordered critical spin chains can be described by a conformally invariant
field theory from which an universal number $c$, the central charge, 
emerges. This central charge, also called conformal anomaly number,
appears in the leading finite size (or finite temperature) correction to the free
energy~\cite{Ian86} as well as in the entanglement entropy
(\ref{eq:2}). The power-law behavior for the spin-spin
correlations functions is also universal with well defined critical
exponents~\cite{LutherPeschel} as well as exact amplitudes~\cite{Amplitude}.

In the case of random critical chains, while the RSRG framework
provides universal critical exponents, the amplitudes of
correlations functions are non-universal numbers~\cite{Fisher-Young}. 
On the other hand, the RSRG treatment for the entanglement
entropy provides the {\it{exact}} prefactor equal to $\frac{\ln
  2}{3}$. This prediction has been checked numerically using exact
numerical diagonalizations on large scale random critical spin
chains. The perfect agreement between exact simulations and the
perturbative RSRG provides, to the best of our knowledge, the first
example of an exact critical amplitude computed within this
technique. It is also interesting to notice that this finding of
${\tilde{c}}=\ln 2 <1$ would be consistent with a generalized
${\tilde{c}}-$theorem built on entanglement concepts for non conformal
random critical points~\cite{Refael04}. Nevertheless, the identificaton of other
physical quantities besides entanglement that are controlled by this
number ${\tilde{c}}$ turns out to be very challenging and more subtle than
using a simple analogy with the clean case. Indeed, since in 
conformally invariant clean systems a non universal
velocity factor appears tied to $c$ in the usual thermodynamic
quantities like the specific heat~\cite{Ian86} or the aforementioned correction to
the free energy, the analogy here breaks down because the velocity of
excitations is not defined anymore in the RSP.

To conclude, we believe that the results presented in this
communication provide a new insight on the random singlet phase as
well as a first visual proof of the large scale effective singlets
formation. The fact that even at the infinite randomness fixed point
the entanglement entropy still scales logarithmically with the
subsystem size provides a non trivial extension of the quantum entanglement
concepts to random quantum critical points.\\

I would like to thank Ian Affleck, Ming-Shyang Chang, Joel Moore, and
Gil Refael 
for
stimulating and interesting discussions, as well as Ingo Peschel and Vladimir
Korepin for correspondence.
I also thank NSERC of Canada for financial support and 
WestGrid for access to computational facilities.




\begin{thebibliography}{}
\bibitem{Nature}A. Osterloh, L. Amico, G. Falci, and R. Fazio, Nature (London) {\bf 4126}, 608 (2002).
\bibitem{Vidal03} G. Vidal, J. I. Latorre, E. Rico, and A. Kitaev, Phys. Rev. Lett. {\bf 90}, 227902 (2003).
\bibitem{HLW94} C. Holzhey, F. Larsen, and F. Wilczek, Nucl. Phys. B {\bf 424} 44 (1994).
\bibitem{Vidal04} J. I. Latorre, E. Ricco, and G. Vidal, Quant. Inf. and Comp. {\bf 4}, 048 (2004).
\bibitem{Korepin04} V. E. Korepin, Phys. Rev. Lett. {\bf 92}, 096402 (2004).
\bibitem{Cardy04} P. Calabrese and J. Cardy, J. Stat. Mech. 06 (2004) 002.
\bibitem{Refael04} G. Refael and J. E. Moore, Phys. Rev. Let. {\bf 93},
  260602 (2004).
\bibitem{Ma} S.-k. Ma, C. Dasgupta, and C.-k. Hu, Phys. Rev. Lett. {\bf{43}}, 1434 (1979); C. Dasgupta and S.-k. Ma, Phys. Rev. B {\bf{22}}, 1305 (1980).

\bibitem{Fisher} D. S. Fisher, Phys. Rev. B {\bf{50}}, 3799 (1994); D. S. Fisher, Phys. Rev. Lett. {\bf{69}}, 534 (1992); D. S. Fisher, Phys. Rev. B {\bf{51}}, 6411 (1995).
\bibitem{note1} To be more precise, this is only true when the
  anisotropy $\Delta\in [-0.5,1]$; see C.~A.~Doty and D.~S.~Fisher,
  Phys. Rev. B {\bf 45}, 2167 (1992).
\bibitem{Peschel1} I.Peschel, M.Kaulke and O.Legeza, Ann. Physik (Leipzig) {\bf 8}, 153 (1999).

\bibitem{Baxter} R. J. Baxter, Exactly Solved Models in Statistical
  Mechanics, Academic Press, London (1982).
\bibitem{Andreas}A. M. La\"uchli: ``Quantum magnetism and strongly
  correlated electrons in low dimension'', PhD Thesis, Swiss Federal
  Institute of Technology, Z\"urich (2002).
\bibitem{Chung01}M.-C. Chung and I. Peschel, Phys. Rev. B {\bf 64},
  064412 (2001; I. Peschel, J. Phys. A {\bf{36}}, L205 (2003); I. Peschel, J. of Statistical Mechanics P06004 (2004).
\bibitem{Peschel2} I. Peschel, J. Phys. A: Math. Gen. {\bf 38}, 4327 (2005).
\bibitem{note2}{Note that we have also
  simulated less disordered systems with narrower distributions but the
conclusions do not change, except for some finite size crossover
phenomena already discussed in Ref.~\cite{comment03} for the
correlation functions.}
\bibitem{comment03} N. Laflorencie and H. Rieger, Phys. Rev. Lett {\bf
  91}, 229701 (2003); N. Laflorencie, H. Rieger, A. W. Sandvik, and
P. Henelius, Phys. Rev. B {\bf 70}, 054430 (2004).
\bibitem{LSM61} E. Lieb, T. Schulz, and D. Mattis, Ann. Phys. (NY)
  {\bf{16}}, 407 (1961).
\bibitem{Henelius98} P.~Henelius and S.~M.~Girvin, Phys. Rev. B {\bf{57}}, 11457 (1998).
\bibitem{LAR} The linear algebra package LAPACK has been used here.
\bibitem{Korepin2}  B.-Q. Jin and V. E. Korepin, J. Stat. Phys. {\bf 116}, 79 (2004).
\bibitem{Masuda04} T. Masuda, A. Zheludev, K. Uchinokura, J.-H. Chung, and S. Park
Phys. Rev. Lett. {\bf 93}, 077206 (2004).
\bibitem{Ian86} H. W. J. Bl\"ote, J. L. Cardy,
and M. P. Nightingale, Phys. Rev. Lett. {\bf 56}, 742 (1986); I. Affleck, 
Phys. Rev. Lett. {\bf 56}, 746 (1986).
\bibitem{LutherPeschel} A. Luther and I. Peschel, Phys. Rev. B {\bf
  12}, 3908 (1975).
\bibitem{Amplitude} I. Affleck, J. Phys. A {\bf 31}, 4573 (1998);
  S. Lukyanov, Phys. Rev. B {\bf 59}, 11163 (1999).
\bibitem{Fisher-Young} D. S. Fisher and A. P. Young, Phys. Rev. B {\bf
  58}, 9131 (1998).
\end{thebibliography}
\end{document}